**SIG-DB: leveraging homomorphic encryption to Securely Interrogate privately held Genomic DataBases**

Alexander J. Titus[1,2,*], Audrey Flower[3], Patrick Hagerty[4], Paul Gamble[3], Charlie Lewis[5], Todd Stavish[3], Kevin P. O'Connell[1], Greg Shipley[5], and Stephanie M. Rogers[1]

[1]B.Next, In-Q-Tel, Arlington, VA, USA

[2]Quantitative Biomedical Sciences, Dartmouth College, Hanover, NH, USA

[3]Lab41, In-Q-Tel, Arlington, VA, USA

[4]CosmiQ, In-Q-Tel, Arlington, VA, USA

[5]Cyber Reboot, In-Q-Tel, Arlington, VA USA

*Corresponding author: bnext@iqt.org

## Abstract

Genomic data are becoming increasingly valuable as we develop methods to utilize the information at scale and gain a greater understanding of how genetic information relates to biological function. Advances in synthetic biology and the decreased cost of sequencing are increasing the amount of privately held genomic data. As the quantity and value of private genomic data grows, so does the incentive to acquire and protect such data, which creates a need to store and process these data securely. We present an algorithm for the Secure Interrogation of Genomic DataBases (SIG-DB). The SIG-DB algorithm enables databases of genomic sequences to be searched with an encrypted query sequence without revealing the query sequence to the Database Owner or any of the database sequences to the Querier. SIG-DB is the first application of its kind to take advantage of locality-sensitive hashing and homomorphic encryption to allow generalized sequence-to-sequence comparisons of genomic data.

**Introduction**

Genomic information is becoming valuable to a variety of applications, including healthcare and industry. Thus, biological data privacy and security is becoming increasingly important, especially as our understanding of how genetic information relates to biological function continues to grow. The genomic sequence of an individual not only provides information of identity, but can reveal family lineage, physical traits, and disease susceptibility. However, there is inherent tension between the Health Insurance Portability and Accountability Act (HIPAA) requirements for patient data privacy, and scientific and/or public health utility of genomic data. For example, researchers who want to study the genetics of disease may wish to compare patient genome databases in a way that allows data mining while complying with HIPAA obligations.

Microbial genomic data is also increasing in quantity and value as microbes are being utilized for industrial, agricultural, and medical purposes. Advances in synthetic biology are creating valuable opportunities in healthcare and industry, where the market in 2016 was $3.9B, and is projected to grow to $11.4B by 2021[1]. In particular, synthetic biology is transforming industrial practices, where microorganisms are exploited to produce commercial products, resulting in customized microbes and unique genomic data. With that comes fierce corporate competition, and the genomic information becomes proprietary. Increasingly, the impact of the human microbiome on health is also an active area of research[2,3]. As we expand our understanding of the microbiome, our personal microbial fingerprint may become part of the data stored in electronic health records[4]. Therefore, we may arrive at a consensus that a person's microbiome is as important to protect as a person's genome. Additionally, genomic information has become an important component of the identification, characterization, and attribution of an unknown microbiological sample, such as an outbreak of a novel infectious agent. To execute the appropriate bioinformatic analyses, known genomic data is required to compare against the sample's genomic data. However, the growing amount of privately held and proprietary genomic information, and the lack of secure computation techniques for sequence-to-sequence comparisons on that data, could hinder the timely genomic analysis of the novel microorganism.

Data in storage can be protected by hardware- or software-based encryption, but sensitive data is vulnerable during processing. Therefore, there is a need for methods, including search and comparison methods, that permit the processing of data while preserving security. Several approaches have been published that demonstrate software-based, secure, multi-party computation for biomedical applications, including Yao's protocol[5], oblivious transfer[6], and many others[7]. Because human genomes are highly conserved, however, much of the existing work is on methods for assessing genomic variation at the SNP[8,9] or causal variant[5] level.

Homomorphic Encryption (HE)[8–10] is a class of methods with properties[11,12] that make sequence-to-sequence comparisons, such as searching polynucleotide sequence databases, amenable to encryption. HE differs from other encryption methods by allowing operations directly on encrypted data without access to a private key. The result remains encrypted and can be revealed to the owner at a later point with their private key. There are two classes of HE: fully HE and partial HE. Fully HE systems allow an arbitrary set of operations (e.g., addition, subtraction, division, and multiplication) to be performed on ciphertext, but currently come at an operationally intractable computational expense. Partial HE systems allow a much smaller subset of operations on ciphertext, with each encryption system developed to support a specific set of operations. The Paillier homomorphic encryption system (PHE)[12] is an additive (+) HE cryptosystem. PHE algorithms can be orders of magnitude faster than fully HE cryptosystems[11,13], and thus are promising for immediate operational applications. To date, there is no published work we are aware of that leverages any version of HE for polynucleotide sequence-to-sequence comparisons.

We have developed an algorithm, Secure Interrogation of Genomic DataBases (SIG-DB), which is designed to enable a two-party exchange. In this exchange, a Querier encrypts a genetic sequence of interest and passes it to a Database Owner. The Database Owner then compares the encrypted sequence to each item in the database and calculates an encrypted similarity score. After comparisons, the encrypted similarity scores are passed back to the Querier, decrypted with the Querier's private key, and interpreted. In this exchange, the query and each element in the database are kept secret while allowing sequence-to-sequence comparisons. We demonstrate the SIG-DB algorithm's performance on microbial genomic data obtained from the National Center for Biotechnology Information (NCBI) Assembly database[14], but see broader applications across any sequence-to-sequence comparison.

**Results**

The SIG-DB algorithm is intended for use between two parties: a Querier and a Database Owner who is willing-but-unable to share his genomic data without having a way to protect it. For example, these parties may be businesses wanting to keep their IP protected while collaborating, hospitals collaborating on research while considering HIPAA requirements, or an investigator and a genomic data company. This protocol requires both the Querier and the Database Owner to be willing participants in the transaction.

There are two major components of the SIG-DB protocol. First, the genomic sequences are converted to a storage efficient data structure. To do so, we chose to use k-mers and locality-sensitive hashing (LSH). The use of k-mers is an established method of breaking up genetic sequences without appreciative data loss; the size of the k-mer can be optimized for a specific application. LSHs are also space efficient data structures that store information as either a 1 or 0 and, unlike Bloom filters, force the data to a preset vector size that is smaller than the original data dimensions. For example, the sequence 'TATCAGA' would represent a 1 in a separate location within the LSH than 'ATCAGAT' (see Figure 1). The size of the LSH is a trade-off between computational runtime and the likelihood of a false match (i.e. hash collision) in the LSH. For our application, the LSH length is set to be five times the length of the longest sequence of all the sequences being compared. Thus, the probability of a hash collision is 18%. This probability can be changed based on user requirements, and specifically, it can be lowered by increasing the size of the LSH (with an increase in runtime), or by using multiple hash functions in conjunction to build a Bloom filter rather than an LSH[15].

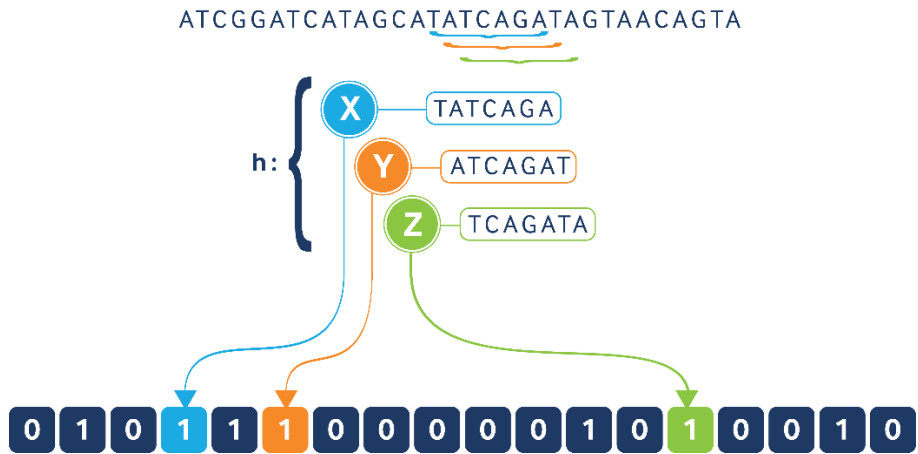

**Figure 1**. Example of a Locality Sensitive Hash (LSH), derived from a DNA sequence. In SIG-DB, the k-mers are created using a sliding window of 1 character, as illustrated, with a sequence of length n resulting in 'n-k' k-mers.

Second, the query LSH data are encrypted, compared to database entries, and resulting scores decrypted. SIG-DB compares encrypted queries against unencrypted genomic sequences in a database while only revealing minimal information about the database to the Querier, and no information about the query to the Database owner. This is achieved by sending the encrypted query to the Database Owner where computation can be carried out behind the database firewall. For each query-database entry LSH comparison, the magnitude of the database entry LSH and an encrypted intersection score are returned to the Querier. From this information, the Querier can calculate similarity scores as the intersection over union (IoU), the intersection over magnitude of query LSH (IoQ), and the intersection over magnitude of database entry LSH (IoD). The full workflow can be seen in Figure 2, outlining the steps in both encrypted and unencrypted space.

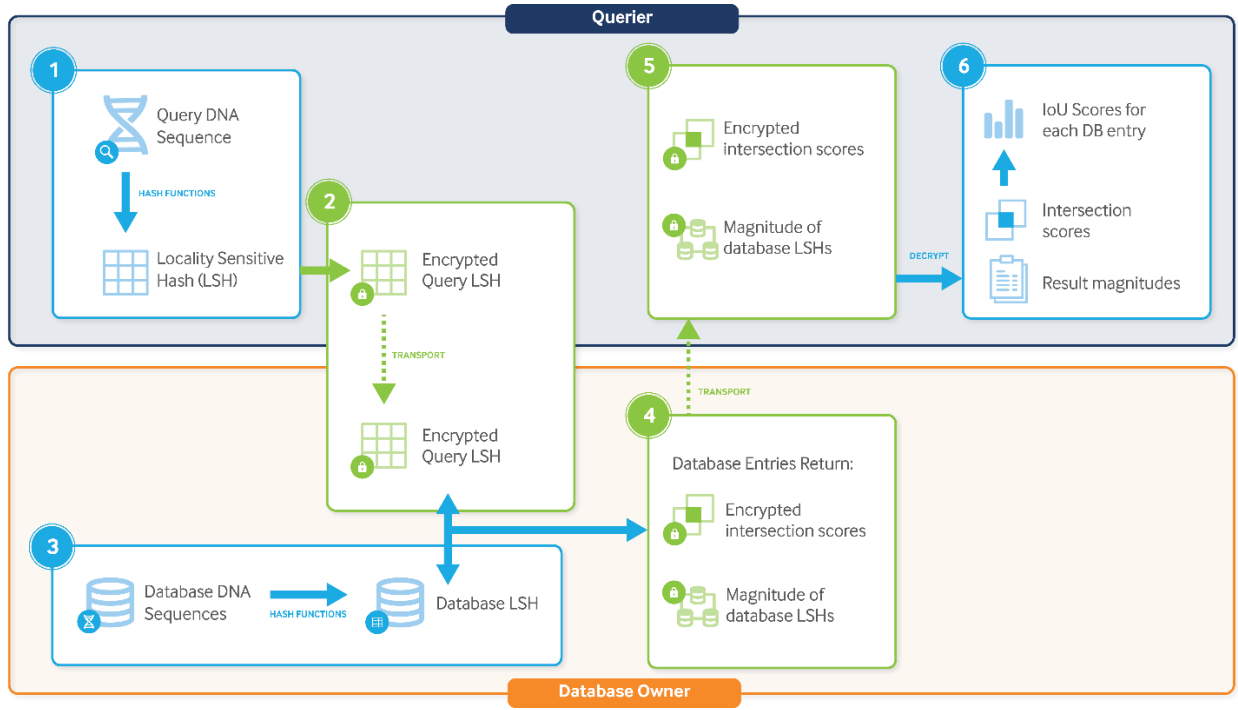

**Figure 2.** SIG-DB protocol for (1) hashing sequence into locality sensitive hash (LSH), (2) encrypting and passing LSH, (3) hashing database elements into LSHs, (4) comparing encrypted query LSH to unencrypted database LSHs, (5) passing scores and decrypting, and (6) calculating IoU, IoD, and IoQ scores. Unencrypted space = blue boxes and encrypted space = green boxes.

### Proportion of sequence mutation & K-mer size

We conducted tuning experiments to identify the effect of k-mer size on algorithm performance and tested the extent to which SIG-DB is robust to sequence mutations, using a dataset of 50 *Escherichia coli* genomes and 50 *Staphylococcus aureus* genomes (from which the query sequences were selected then compared against). The k-mer length was tested at k={8, 16, and 32}. Uniformly distributed, random mutations were introduced to query sequences *in silico* in 5% increments, ranging from 0%-100%. Using a k-mer length of 8, the SIG-DB algorithm correctly returned the sequence of interest as the highest IoU score for sequences with mutation rates from 0-45% for *E. coli* and 0-35% for *S. aureus* (Table 1). The mutation rate tolerance decreases as the k-mer size increases (Figure 3). We found that the method correctly identified the presence of similar sequences in the subject database to a greater level of divergence when k was set to 8. Therefore, we set k=8 for all subsequent work.

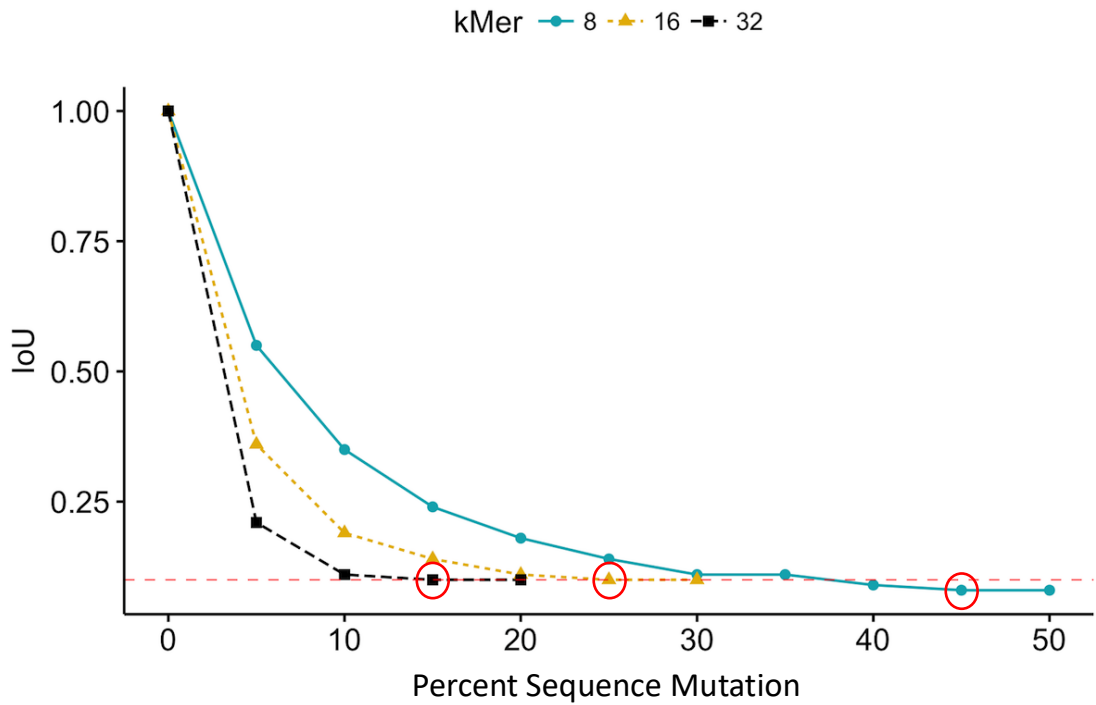

**Figure 3**. Intersection over Union (IoU) scores for k-mer = {8, 16, 32} with random query sequence mutation rates from 0-100%. K-mer = 8 showed best performance, with correct identification of sequence of interest up to mutation rates of 45%. The red circles indicate the largest mutation rate for each k-mer size that returned the correct result as the highest IoU value. The horizontal red line represents IoU = 0.1.

The similarity scores are drastically reduced with minor point mutations because a single mutation affects every *k*-mer associated with it. For example, the sequence TCGATCGATCGATCGA with a point mutation at the highlighted "A" represents a mutation in the 8-mers TCGATCGA, CGATCGAT, GATCGATC, ATCGATCG, TCGATCGA, CGATCGAT, GATCGATC, ATCGATCG. For this reason, shorter *k*-mers are more robust to data degradation and thus lead to better SIG-DB algorithm performance. However, decreasing k-mer size comes at a computational cost; therefore, algorithm performance is a trade-off between k-mer size and run-time.

**Location of sequence mutation**

The previous mutation tests assumed a random distribution of mutations throughout the genomic sequence, representing single base pair discrepancies between a query and a database sequence. To test the effect of mutations localized to one-half of a sequence with the other half a perfect match to an element of the database (i.e. sequences that overlap by 50%), we assessed the performance of the algorithm with 0-50% mutation.

Using a data set of 50 *E. coli* genomes, an *E. coli* query sequence, and a *k*-mer length of *k=8*, the SIG-DB algorithm correctly returned the query sequence of interest as the highest IoU score for sequences with mutation rates from 0% (IoU = 1.0) to 45% (IoU = 0.41). A 0% mutation represents two identical sequences, and a 50% mutation represents two sequences that overlap and match by exactly 50% (Table 2). The similarity scores had a less dramatic decline when mutations are localized to a section of the sequence, compared to the sharp decline that occurred when the mutations are dispersed throughout the entire sequence. This observation is expected, because fewer k-mers are affected by the mutations when they are localized to one half of the sequence (Supplemental Figure S1).

**Query runtime relative to DB size**

To ensure a secure query does not reveal information about the query through execution time, the total run-time must scale linearly with the size of the database being searched. This prevents the point at which the algorithm stops from revealing which database entry met the appropriate search criteria. To test the linear scalability of the SIG-DB algorithm, queries were executed against databases containing an increasing number of sequences for four different query sizes. Sequences of length 100, 1000, 10000, and 20000 base pairs were queried against databases with 7, 70, 700, 7078, and 75958 entries, respectively, and demonstrated a linear relationship between query execution time and database size (Table 3).

**Length of query relative to DB entries**

Biological sequences inherently are quite variable in size, with genomes ranging from thousands (viruses) to billions (plants and mammals) of base pairs, depending on the species. Therefore, it is likely the query sequence will not be the same size as the database entries, and that the database entries will vary in length. We included 3 similarity metrics (IoU, IoQ, and IoD) in the output to account for sequence size variation. To test the robustness of SIG-DB and the similarity metrics for queries and database entries of varying relative lengths, using a data set of 50 *E. coli* genomes and an *E. coli* query sequence, we compared performance on sequences that ranged in relative length from 1:1 to 4:1/1:4 (Q:DB, respectively). The sequence lengths tested varied from 5,000bp to 20,000bp, which represent the typical size ranges for viral genomes and bacterial genes. The k-mer size was held constant at k=8, and no mutations were introduced. For all ratios tested, SIG-DB returned the correct sequence as the highest IoU (Table 4). When the database sequences were held constant at 20,000bp and the query sequence reduced to smaller sizes, the IoQ=1.0 for every test. When flipped, the IoD=1.0 for every test. This performed as expected, because for IoQ (or IoD), only the length of the query (or database entry, respectively) is factored into the similarity score calculation. Therefore, the scoring would be a perfect match for all ratios tested (Supplemental Figure S2).

**Correctness**

To test the robustness of the algorithm to return the correct results, we conceptualized the algorithm as a classifier and tested the classification accuracy by setting a similarity score threshold, above which the algorithm can be thought of as reporting that a particular database entry was similar to the query. We used Average Nucleotide Identity (ANI), a measure of pairwise bidirectional nucleotide similarity, as the ground truth similarity measurement for each query-database entry comparison, with ANI thresholds of 95% and 99%[16,17].

The algorithm performed well across all database entry length sets and under both ANI thresholds (AUC values range from 0.97 to 1.00). In fact, with an ANI threshold of 0.95 (a slightly more 'permissive' gold standard), the algorithm was essentially a perfect classifier

(AUC=1.00 for all length sets) - assigning similarity scores to database entry-query pairs in exactly the same order as their ANI score rankings (Supplemental Figures S3, S4).

**Practicality**

One major concern for using homomorphic encryption techniques for operational purposes is the computational expense. Total execution time for SIG-DB ranges from a few seconds to approximately 1300 seconds when using a single CPU core to process a single query against a database of 100 entries with sequence lengths of 100 and 20,000 bases, respectively (Supplemental Figure S5). Parallelization helps performance but has diminishing returns for more than 16 CPU cores. For example, with a 20,000 bp query sequence length and a database of 82,992 sequences, the execution time goes from 51 minutes with one core to 21 minutes with 16 cores - a reduction of more than 50%. However, when looking only at scoring time, performance is worse when the algorithm is parallelized. This observation is likely related to the size of the query being scored and how efficient it is, or is not, to divide the work over several CPUs at the same time, versus running the calculation on a single CPU without that overhead. Additionally, query comparison time and similarity scoring time are the most time intensive steps, and they appear to be equal even as parameters change (Supplemental Figure S6).

The linear scalability of execution time is an important requirement of HE algorithms. If this does not occur, then one or more parties may infer information about the encrypted data from the point in which computation ends. For example, if searching for an exact match in a database, if the algorithm terminates before searching all elements of the database, then the database holder can infer that the last element searched was an exact match for the query. Our algorithm searches every element of a database and was shown to scale linearly as a function of database size (factor of 8 difference between total execution time of tests ran with 100 and 1000 database entries; Supplemental S7).

**Discussion**

We developed the secure interrogation of genomic databases (SIG-DB) algorithm as a proof-of-concept for a suitable, technologically feasible approach to compare a genomic sequence of interest to a privately-held database and produce an indication of similarity between the sequence of interest and each database entry. This approach could enable bioinformatics practitioners or investigators to leverage privately held information in a secure way – a capability that is non-existent today for genomic sequence comparisons. This tool could expedite the timeliness of queries and gain meaningful results in critical timeframes for infectious disease event response decisions. This is important, as many bioinformatics practitioners are not security experts, and thus are less likely to develop applications in this space.

The overall security of an encrypted data manipulation algorithm is fundamental to its success. The SIG-DB algorithm leverages established homomorphic encryption schemes to ensure that no outside (non-participating) party can learn the details of specific queries; only information about the query request patterns can be learned. It also ensures that no outside party can view the database records directly. Only metadata is exchanged in the form of encrypted LSH magnitudes, and thus individual data entries are never exposed outside the owners' own security systems.

From the metadata returned, the Querier will learn the number of entries in the database. When combined with the DB LSH magnitudes and the similarity scores returned, the Querier could extrapolate the diversity within the database. Additionally, if the Querier repeated SIG-DB multiple times against the same database using the same query sequence, the Querier could learn about any database modifications that may have occurred. This risk is limited, however, since the execution is controlled by the Database Owner, who retains the right to refuse repeat executions.

A fully homomorphic encryption (FHE) approach, such as Microsoft SEAL, reduces the amount of information leaked by not needing to reveal the magnitudes of Database LSHs (unlike PHE), and therefore does not reveal the length of the database entries. This is because FHE allows for division and multiplication, allowing the algorithm to calculate the full similarity score (while keeping it encrypted) before passing the information out of the database. However, FHE comes with an increased computational cost (an FHE implementation of SIG-DB required 2.25 times as long to execute as the PHE implementation; unpublished results), which currently limits its operational utility.

We developed this tool with the initial intention of it being used for microbial genomic sequence comparisons, especially when a user needs to quickly identify a source of information relevant to the query sequence, such as an infectious disease outbreak or intentional release of a bioweapons agent. However, there are a number of applications that it could be used for including human genomics, healthcare, organizational collaborations, and more. SIG-DB can be optimized for a particular application by modifying the parameters (k-mer size, LSH size, hash collision rate) based on the user's tolerance for mutations, size of database, and computational resources. First, the size of the k-mer should be chosen to account for the degree of variability that is captured. Our results indicate that a smaller k-mer is robust to mutations and that tolerance is reduced by increasing the size of the k-mer. The operator will need to decide how much variation should be tolerated and establish their k-mer size accordingly. Second, the current implementation reports three scoring metrics: IoU, IoD, and IoQ. These were included to account for potential variations in sequence size and locations of mutations. Lastly, computational resources should be considered and optimized for the given situation. The process of sequence comparisons and decryption will be the most computationally burdensome given the operations over every database entry. This should be considered by the query holder when establishing computational infrastructure. Of note, the SIG-DB prototype was developed in Python, and thus runs with the efficiency of a Python program. Extensions of this work should include implementing the SIG-DB algorithm in C++ to allow for faster runtime.

Overall, the SIG-DB algorithm provides a method for a secure multi-party exchange of information. The outcome of this protocol is a similarity score, which provides an indication of the relative similarity between a query and the interrogated database. Should the user determine there is adequate similarity, the next steps would be based on his needs and processes. We see use cases within organizations where parties are willing and open to participate in the exchange, but where security of information even within the organization is paramount, such as in the case of HIPAA data in healthcare. The potential versatility of this protocol creates an even greater opportunity for impact within the broader scientific community.

**Methods**

**Data**

We accessed and downloaded available bacterial genome data in FASTA format from the National Center for Biotechnology Information (NCBI) Assembly RefSeq[14] database (DB), resulting in a total dataset of 75,958 samples. The dataset contained 244 gigabytes of complete genome, chromosome, scaffold, and contig sequences for all available bacterial species. For algorithm testing, we used *E. coli* ($n = 7,078$ genomes) and *S. aureus* genomes ($n = 50$ genomes) as well as the full dataset of all available species genomes ($n = 75,958$).

**Query preparation**

To build the LSH with the appropriate false positive rate, the maximum length of a sequence in the database must be obtained from the Database Owner. From this implementation, the LSH is constructed to have a 5:1 ratio of available space to filled locations. For testing purposes, sequences with maximum length of 20,000 base pairs were used, thus LSHs were initialized to have 100,000 available hash locations. For sequences longer than 20,000 base pairs, the first 20,000 bases were used, and sequences shorter than 20,000 base pairs were used in their entirety. Sequence *k*-mers were hashed into the LSH using the Python *hash* function.

Homomorphic encryption was implemented with the Python package *phe*[18], implementing the Paillier additive homomorphic encryption system (PHE)[12]. Under this system, a public key/private key pair is generated and each element of an LSH is encrypted using the public key.

**Database searching**

An LSH is created for each database entry using the LSH Constructor, provided by the Querier. The Database Owner executes the comparison between each DB entry LSH and the encrypted Query LSH using the comparison executable provided by the Querier. For each comparison, an encrypted intersection score is calculated by the sum of all the encrypted Query LSH hash locations corresponding to a filled hash entry in the DB entry's LSH. The magnitude of the DB entry LSH is calculated as the sum of hashes in the LSH. The pair {encrypted intersection score, DB entry LSH magnitude} are then returned to the Querier for evaluation.

**Similarity metric calculations**

To assess the similarity between the query and the entries in the database, the returned encrypted intersection scores are decrypted using the private key. The intersection over union (IoU) score is calculated using the unencrypted intersection score, the query LSH magnitude, and the DB entry LSH magnitude:

$$IoU = \frac{Intersection}{|Query\ LSH| + |DB\ entry\ LSH| - Intersection}$$

Additionally, to account for scenarios with a query sequence shorter than the database sequences, the intersection over length of query (IoQ) is calculated using the unencrypted intersection score and the query LSH magnitude:

$$IoQ = \frac{Intersection}{|Query\ LSH|}$$

Finally, to account for scenarios with a query sequence longer than the database sequences, the intersection over DB entry LSH magnitude (IoD) is calculated using the unencrypted intersection score and the DB entry LSH magnitude:

$$IoD = \frac{Intersection}{|DB\ entry\ LSH|}$$

From the IoU, IoQ, and IoD, the Querier can evaluate the highest similarity and the overall similarity of database entries to the query sequence.

**SIG-DB Algorithm**

The SIG-DB algorithm is a two-party exchange for sequence-to-sequence comparisons between an encrypted query and elements of a database. The steps of the algorithm are as follows:

1. Querier: From Database Owner, determine largest sequence length in database

2. Querier: Hash query sequence $k$-mers into a locality-sensitive hash (LSH) with a specified acceptable false positive rate based on largest entry sequence length

3. Querier: Generate a public/private key pair using the Paillier cryptosystem

4. Querier: Encrypt each element in the query LSH using the Paillier public key

5. Querier: Pass the encrypted query LSH, the LSH constructor with hash function, and a processing script to the Database Owner

6. Database Owner: Hash each database sequence entry into an LSH of matching dimension to the query LSH using the original hash function, provided by the Querier.

7. Database Owner: For every database entry LSH :

    1. Calculate encrypted intersection score by the sum of all elements in the encrypted query LSH that correspond to a filled entry in the DB LSH

    2. Calculate the magnitude of the DB LSH

8. Database Owner: return all pairs (DB LSH magnitude and encrypted intersection score) to the Querier

9. Querier: decrypt the encrypted intersection scores using the Paillier private key

10. Querier: For each query-database entry pair, calculate the intersection over union (IoU), intersection over query magnitude (IoQ), and intersection over DB LSH magnitude (IoD).

**SIG-DB Algorithm testing**

Our algorithm testing was conducted on an NVIDIA Titan X GPU with 64 GB of memory. The initial implementation of SIG-DB uses 48 parallel processes to encrypt the LSH and generate the reported similarity metrics.

We tested the algorithm using the data downloaded from the NCBI Assembly database. We used an LSH with 500-100,000 available locations for hashing relative to standardized sequence lengths of 100-20,000 base pairs (always in 5:1 ratio, respectively). These lengths were chosen to represent a range of sequences from raw sequence read to small genome sizes. *E. coli* genomes range from between 4,500,000-6,000,000 bases, but for efficiency, we used a reduced fraction of the full genome as the first *n* base pairs of the sequence (seq[1:n]). The algorithm scales to encompass full genome size sequences, but appropriate adjustments to LSH size must be made.

### *In silico* sequence mutations

We introduced mutations into wild-type query sequences randomly across queries of various lengths. These mutations were intended to represent point mutations in the sequence as well as loss of data quality due to sequencing errors. The point mutations were conducted by changing the selected base pair to either one of the remaining three base pairs, or to an 'X' representing a data error. These mutations were introduced at random distributions across the sequence, ranging from no mutations (0%) to full sequence mutations (100%).

### Correctness

A dataset was created from two *E. coli* genomes (K-12 MG1655 and O157:H7, taken from the RefSeq database). 500 non-overlapping database entries were created by sampling at random from the full genomes for each of three lengths: 1000, 2000, and 3000 bp. From each database entry, a 1000 bp query sequence was selected at random. (Queries taken from a database entry of a given length were used only in searches against the database set from which they were generated.) In order to have sufficient coverage of the desired genetic diversity, the selected query sequences were duplicated twice and the duplicates randomly mutated, where each base pair in the two duplicates had a 5% or 10% chance (respectively) of being switched to an alternate base. In total, this produced 1,500 queries for each length-set of database entries: 500 wild-type, 500 samples with 5% mutation, and 500 samples with 10% mutation.

ANI and SIG-DB similarity scores were computed for each database entry-query pair within a given length-set. For the SIG-DB score, the maximum of the IoU, IoQ, or IoD was taken to account for sequence:query length differences. A Receiver Operator Characteristic (ROC) curve was generated to determine the tradeoff between true positive rate (TPR) and false positive rate (FPR). Multiple classification thresholds were defined for the SIG-DB similarity scores such that all sequences with similarity ≥ threshold were considered "relevant". The thresholds were set in an increasing manner, starting with Similarity = 0 representing a threshold that counted all sequences as "relevant". Each subsequence threshold was then set as the next highest similarity score in the result set. For example, if five sequences had similarity scores of [0.2, 0.33, 0.33, 0.75, 0.87] then each of the unique scores would be set as a classification threshold, resulting in the following counts of the number of "relevant" sequences at each threshold: [5, 4, 2, 1]. For each new threshold, the TPR and FPR were calculated, then plotted as a ROC curve.

**Practicality**

To assess the practicality of using SIG-DB in operational applications, we performed a series of tests to understand the amount of time required to run the full operation on a single CPU core, the time savings that could be achieved using a parallelized approach, and the breakout of time requirements by step in the protocol (encryption time, query time, and scoring time). All bacterial sequences in the RefSeq database were used to perform these tests. The database records have a minimum length of 337 bases, a maximum length of 12,106,419 bases, and an average length of 2,756 bases. Heat maps were generated from test runs that evaluate 3 different variables: (1) number of cores used for computation (1 - 48 cores), (2) query length (100 – 20,000 bases), and (3) number of entries in the database (100, 1000, 5000, and 10000 FASTA files; or 10,145 sequences, 89,992 sequences, 425,772 sequences, and 860,984 sequences, respectively).

**Code/Data availability:**

All code and documentation is available on the B.Next GitHub page, located at https://github.com/BNext-IQT/GEMstone. All sequence data were obtained from https://www.ncbi.nlm.nih.gov/assembly. We accessed and downloaded all sequence data used in this study on October 1st, 2017.

**Acknowledgements**

The authors would like to thank Lisa Porter for her helpful discussions during development of this project and her critical reading of the manuscript; and Nat Puffer and Justin Wilder for their feedback and discussion on encryption and data security.

**Author contributions**

AJT developed the algorithm, conducted analyses, and wrote the manuscript. AF developed the initial algorithm and conducted preliminary analyses. PH, GS, KPO'C, and TS conceived of the project and developed the research plan. PG performed the ROC curve analysis. CL developed and executed the practicality analysis. SR conceived of the project, developed the research plan, and wrote the manuscript.

**Competing financial interests**

All authors certify there are no competing financial interests related to this work.

## Tables

Table 1. SIG-DB algorithm performance based on proportion of sequence randomly mutated in *E. coli* and *S. aureus* (DB = 50 seqs, LSH = 100k bases, Q = 20K bases, E = 20K bases)

| Species | K-mer size | Percent mutation** | Best | | |
|---|---|---|---|---|---|
| | | | IoU | IoQ | IoR |
| *E. coli* | 8 | 0 | 1.00 | 1.00 | 1.00 |
| *E. coli* | 8 | 10 | 0.35 | 0.49 | 0.56 |
| *E. coli* | 8 | 20 | 0.18 | 0.28 | 0.33 |
| *E. coli* | 8 | 30 | 0.11 | 0.19 | 0.22 |
| *E. coli* | 8 | 40 | 0.09 | 0.16 | 0.18 |
| *E. coli* | 8 | 50 | 0.08 | 0.15 | 0.14 |
| *E. coli* | 16 | 0 | 1.00 | 1.00 | 1.00 |
| *E. coli* | 16 | 10 | 0.19 | 0.32 | 0.32 |
| *E. coli* | 16 | 20 | 0.11 | 0.20 | 0.20 |
| *E. coli* | 16 | 30 | 0.10 | 0.19 | 0.19 |
| *E. coli* | 32 | 0 | 1.00 | 1.00 | 1.00 |
| *E. coli* | 32 | 10 | 0.11 | 0.20 | 0.20 |
| *E. coli* | 32 | 20 | 0.10 | 0.19 | 0.19 |
| *S. aureus* | 8 | 0 | 1.00 | 1.00 | 1.00 |
| *S. aureus* | 8 | 10 | 0.35 | 0.46 | 0.58 |
| *S. aureus* | 8 | 20 | 0.18 | 0.26 | 0.35 |
| *S. aureus* | 8 | 30 | 0.11 | 0.17 | 0.23 |
| *S. aureus* | 8 | 40 | 0.08 | 0.15 | 0.16 |
| *S. aureus* | 16 | 0 | 1.00 | 1.00 | 1.00 |
| *S. aureus* | 16 | 10 | 0.18 | 0.31 | 0.31 |
| *S. aureus* | 16 | 20 | 0.11 | 0.21 | 0.21 |
| *S. aureus* | 16 | 30 | 0.10 | 0.19 | 0.19 |
| *S. aureus* | 50 | 0 | 1.00 | 1.00 | 1.00 |
| *S. aureus* | 50 | 10 | 0.10 | 0.19 | 0.19 |

Table 2. SIG-DB algorithm performance based on mutations localized to one half of total sequence (DB = 50 seqs, LSH = 100k bases, Q = 20K bases, E = 20K bases)

| Species | K-mer size | Percent mutation** | Best | | |
|---|---|---|---|---|---|
| | | | IoU | IoQ | IoR |
| E. coli | 8 | 5 | 0.73 | 0.82 | 0.87 |
| E. coli | 8 | 10 | 0.60 | 0.71 | 0.78 |
| E. coli | 8 | 15 | 0.52 | 0.64 | 0.72 |
| E. coli | 8 | 20 | 0.46 | 0.59 | 0.67 |
| E. coli | 8 | 25 | 0.44 | 0.57 | 0.65 |
| E. coli | 8 | 30 | 0.42 | 0.56 | 0.64 |
| E. coli | 8 | 35 | 0.41 | 0.55 | 0.62 |
| E. coli | 8 | 40 | 0.41 | 0.55 | 0.61 |
| E. coli | 8 | 45 | 0.41 | 0.56 | 0.61 |
| E. coli | 8 | 50 | 0.42 | 0.57 | 0.61 |

Table 3. SIG-DB runtime for increasing database sizes (Mut = 0, IoU = 1.0, K = 8)

| Species | Database size | LSH Size | Query size & DB element size | | Run time (min) |
| --- | --- | --- | --- | --- | --- |
| E. coli | 7 | 500 | 100 | 100 | 0.03 |
| E. coli | 70 | 500 | 100 | 100 | 0.03 |
| E. coli | 700 | 500 | 100 | 100 | 0.07 |
| E. coli | 7,078 | 500 | 100 | 100 | 0.45 |
| E. coli | 75,958 | 500 | 100 | 100 | 4.58 |
| E. coli | 7 | 5,000 | 1,000 | 1,000 | 0.09 |
| E. coli | 70 | 5,000 | 1,000 | 1,000 | 0.10 |
| E. coli | 700 | 5,000 | 1,000 | 1,000 | 0.65 |
| E. coli | 7,078 | 5,000 | 1,000 | 1,000 | 5.44 |
| E. coli | 75,958 | 5,000 | 1,000 | 1,000 | 101.64 |
| E. coli | 7 | 50,000 | 10,000 | 10,000 | 0.43 |
| E. coli | 70 | 50,000 | 10,000 | 10,000 | 1.25 |
| E. coli | 700 | 50,000 | 10,000 | 10,000 | 8.35 |
| E. coli | 7,078 | 50,000 | 10,000 | 10,000 | 86.85 |
| E. coli | 75,958 | 50,000 | 10,000 | 10,000 | 993.82 |
| E. coli | 7 | 100,000 | 20,000 | 20,000 | 0.85 |
| E. coli | 70 | 100,000 | 20,000 | 20,000 | 2.69 |
| E. coli | 700 | 100,000 | 20,000 | 20,000 | 18.22 |
| E. coli | 7,078 | 100,000 | 20,000 | 20,000 | 185.1 |
| E. coli | 75,958 | 100,000 | 20,000 | 20,000 | 2215.25 |

*7,078 = Total number of E. coli genomic samples

**75,958 = Total number of all available genomic samples

Table 4. SIG-DB algorithm performance based on relative sizes of query and database elements (K = 8, DB = 50 seqs, LSH = 100k bases, Mut = 0)

| Species | Query size & DB element size | | Best | | |
|---|---|---|---|---|---|
| | | | IoU | IoQ | IoR |
| E. coli | 20,000 | 20,000 | 1.0 | 1.0 | 1.0 |
| E. coli | 15,000 | 20,000 | 0.80 | 1.0 | 0.80 |
| E. coli | 10,000 | 20,000 | 0.58 | 1.0 | 0.58 |
| E. coli | 5,000 | 20,000 | 0.31 | 1.0 | 0.31 |
| E. coli | 20,000 | 15,000 | 0.80 | 0.80 | 1.0 |
| E. coli | 20,000 | 10,000 | 0.58 | 0.58 | 1.0 |
| E. coli | 20,000 | 5,000 | 0.31 | 0.31 | 1.0 |

**Supplemental Tables:**

Table S1. ANI Data statistics describing the distribution of the ANI scores for each database element length dataset.

| Database element Length | Min | First Quartile | Mean | Third Quartile | Max | STD |
|---|---|---|---|---|---|---|
| 1000 | 0.00 | 0.34 | 0.41 | 0.53 | 1.00 | 0.09 |
| 2000 | 0.00 | 0.36 | 0.42 | 0.51 | 1.00 | 0.12 |
| 3000 | 0.00 | 0.29 | 0.46 | 0.59 | 1.00 | 0.19 |

**Supplemental Figures:**

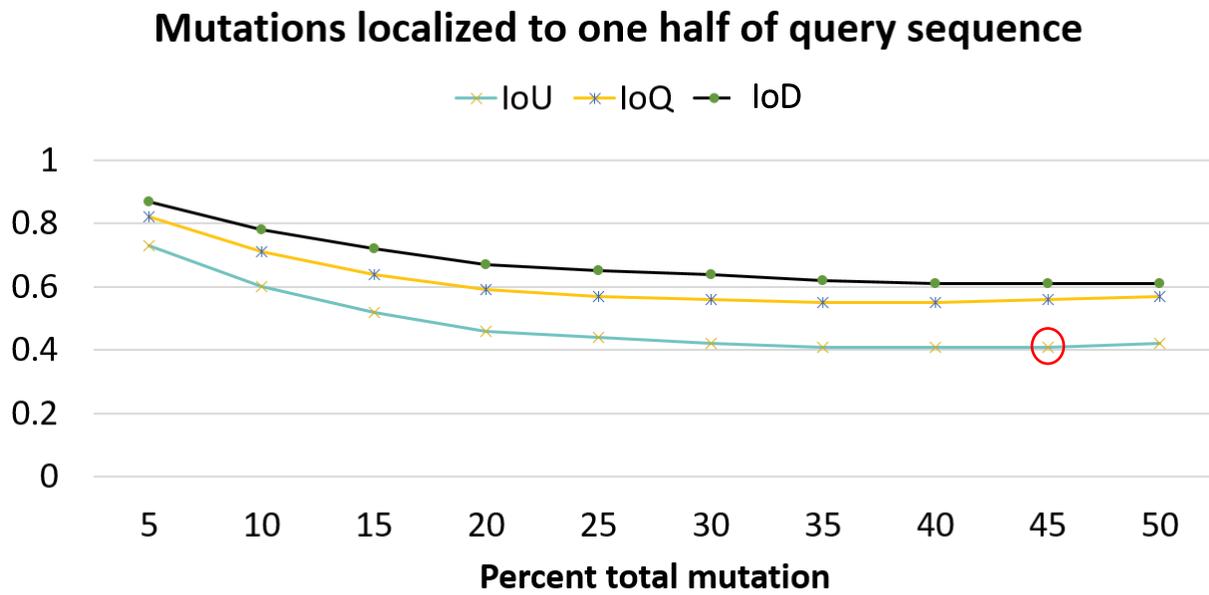

**Figure S1**: Similarity scores (IoU, IoQ, and IoD) for k-mer = {8} with random sequence mutation localized to one half of the sequence at rates from 0-100%. The red circle represented the largest mutation rate that correctly returned the sequence of interest as the highest IoU score.

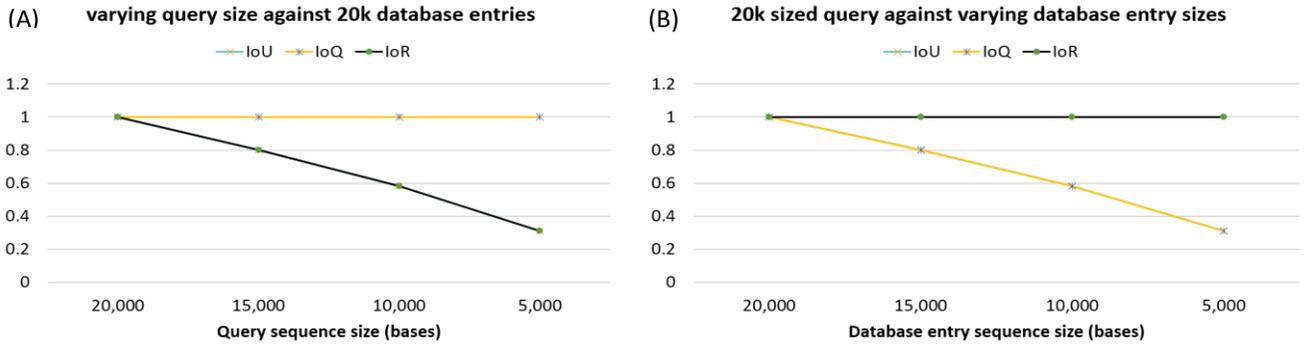

**Figure S2**: Similarity scores for varying sizes of query and DB entries. (A) DB held constant at 20,000 bases while query varied from 5,000-20,000 bases. (B) Query held constant at 20,000 bases while DB varied from 5,000-20,000 bases.

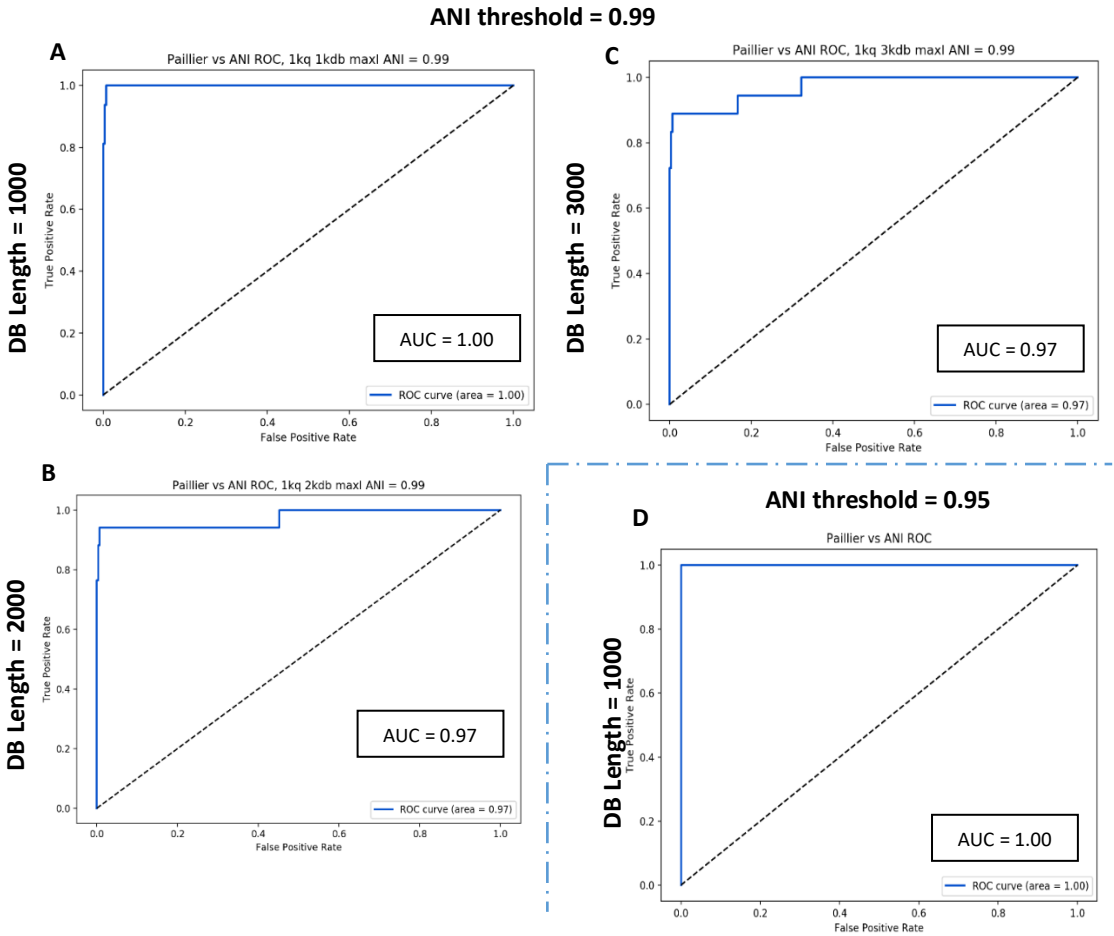

**Figure S3**: Receiver Operator Characteristic (ROC) curves for Paillier-SIG-DB. Query length =1000 bases and ANI threshold of 0.99 for A-C; 0.95 for D. (A) DB entry length=1000 bp with a total of 750,000 query to database entry comparisons and 1,515 of those with ANI scores above the ANI threshold (true positives) (B) DB entry length=2000 bp with a total of 750,000 query to database entry comparisons and 1,549 of those with ANI scores above the ANI threshold (C) DB entry length=3000 bp with a total of 750,000 query to database entry comparisons and 1,708 of those with ANI scores above the ANI threshold (D) DB entry length=1000bp with a total of 750,000 query to database entry comparisons and 1,877 of those with ANI scores above the ANI threshold (representative of all ROC curves generated using an ANI threshold=0.95).

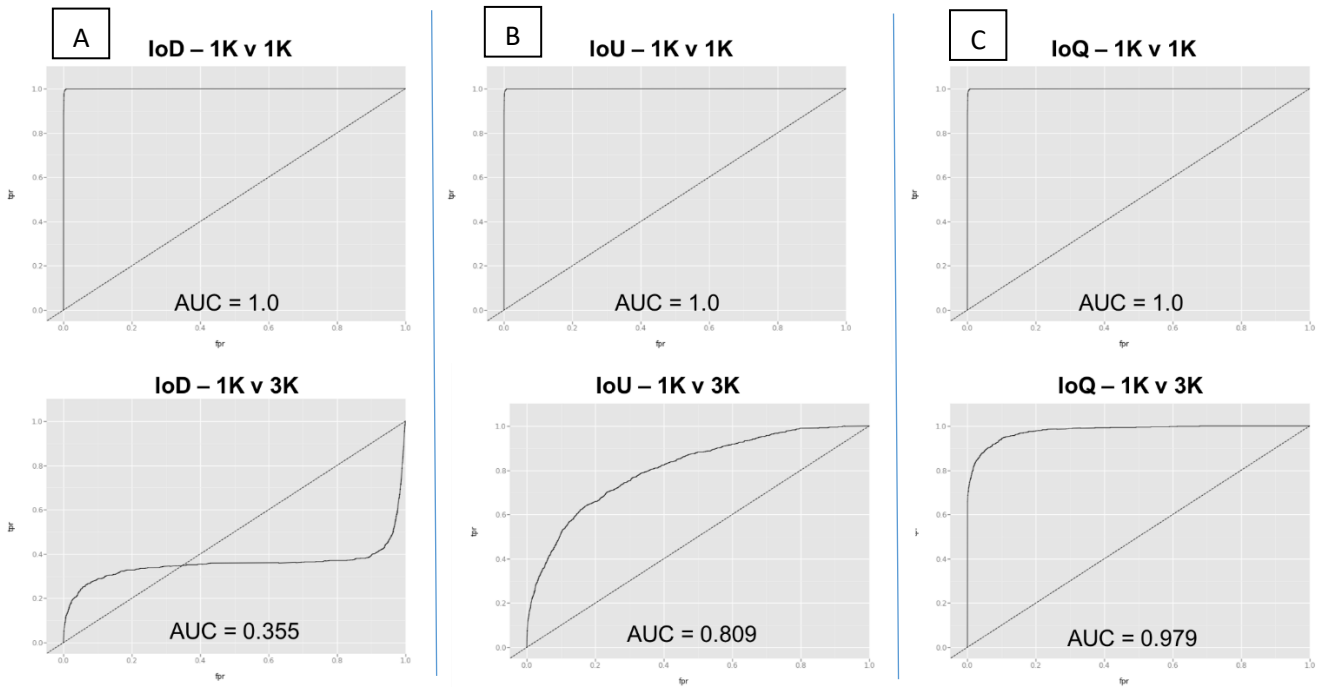

**Figure S4**: Receiver Operator Characteristic (ROC) curves for Paillier-SIG-DB based on each similarity score – (A) IoD, (B) IoU, and (C) IoQ. Query length =1000 bases and ANI threshold of 0.99. The DB entry length was 1000 bp (top row) with a total of 750,000 query to database entry comparisons and 1,515 of those with ANI scores above the ANI threshold (true positives) and 3000 bp (bottom row) with DB entry length=3000 bp with a total of 750,000 query to database entry comparisons and 1,708 of those with ANI scores above the ANI threshold. Using IoD as the similarity score when a query is one-third the size of the database entries results in an odd ROC curve representing very poor performance, as shown in (A) bottom row. One possible explanation is that the denominator is larger than the possible intersection space, and as such, we may be washing out the small differences between the intersections of positive and negative examples. In a dataset of only 0.2% positive samples, this leads to broad misclassification. This phenomenon is exactly why we included IoD, IoU, and IoQ as our scoring metrics.

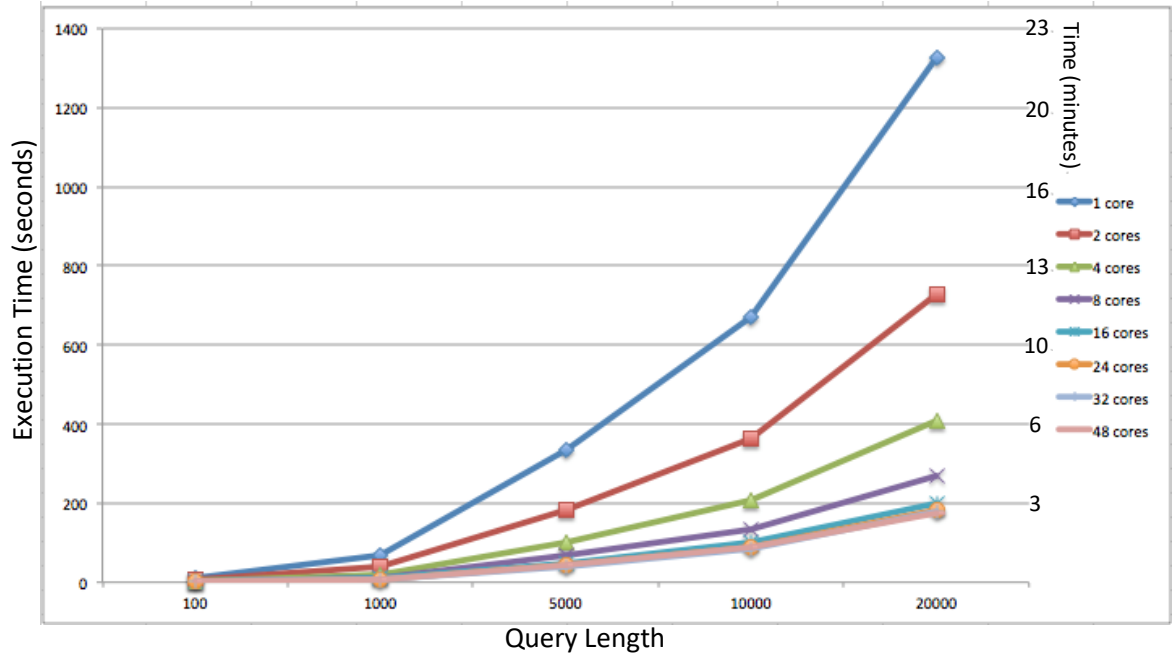

**Figure S5**. SIG-DB Execution time for a query against a database of 100 entries (10,145 sequences) using various numbers of cores for computation.

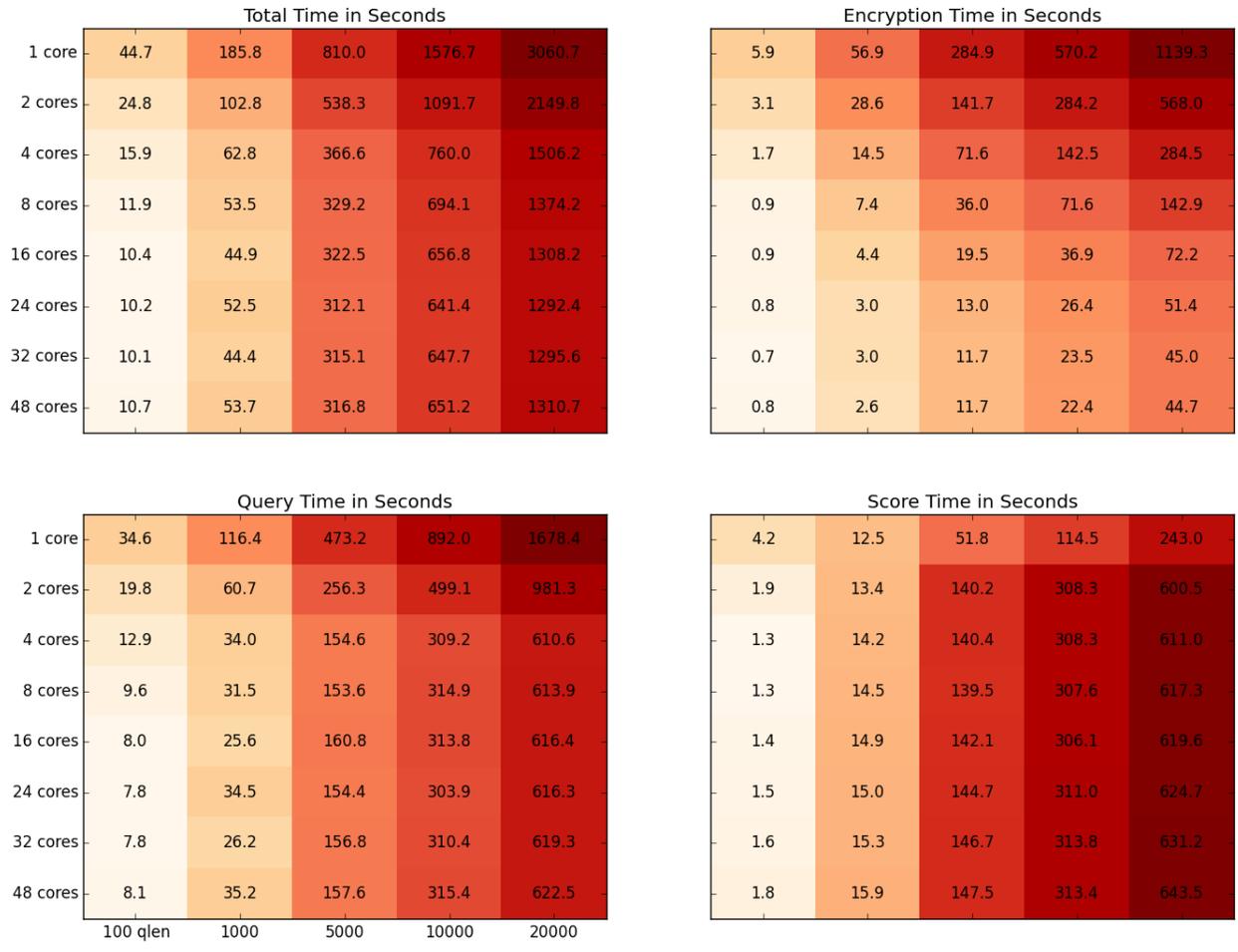

**Figure S6**: Heat Map showing time of execution for SIG-DB, comparing a single query of 'qlen' length to a database containing 1000 entries (82,992 sequences). The trends in this heat map are representative of all scenarios tested.

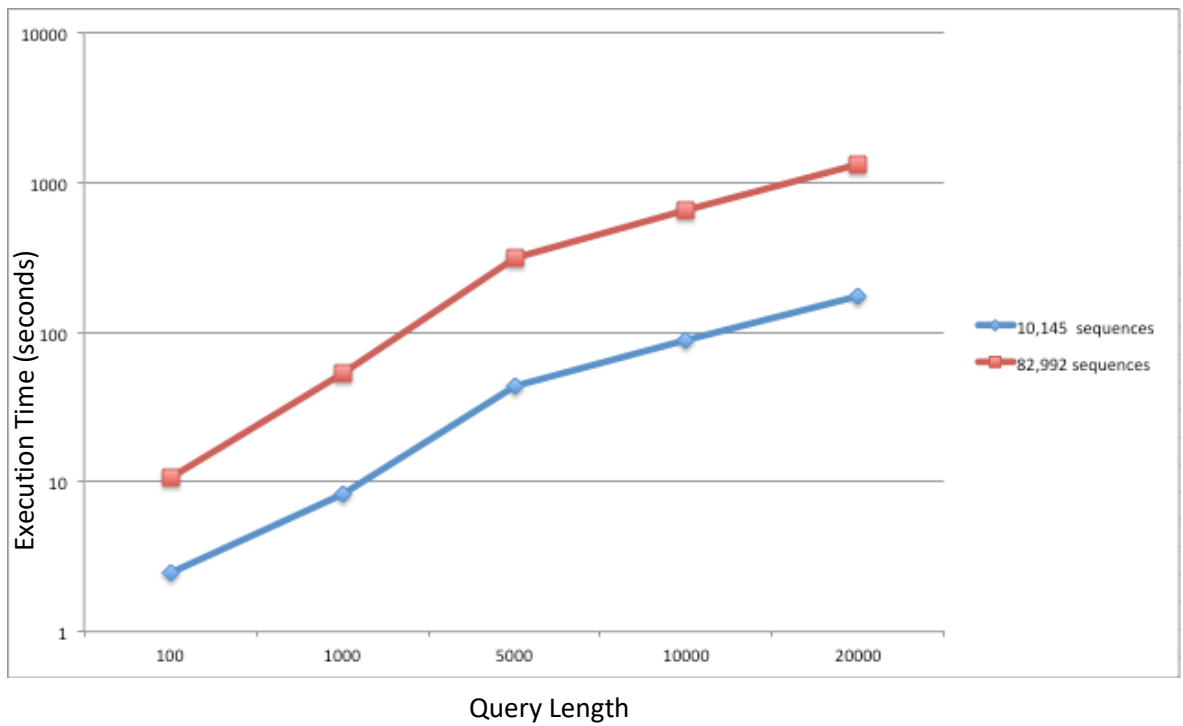

**Figure S7**: SIG-DB execution time for 100 and 1000 database entries compared to different query lengths; both runs performed with 48 cores. Note: y-axis is in log-scale.